\documentclass[aps,prb,twocolumn,superscriptaddress,amsmath,amssymb,floatfix]{revtex4}
\usepackage{graphicx}
\usepackage{dcolumn}
\usepackage{bm}
\usepackage{epsfig}
\def\prb{Phys. Rev. B}
\def\prl{Phys. Rev. Lett.}

\def\pr{Phys. Rev.}
\def\be{\begin{equation}}
\def\ee{\end{equation}}
\def\ba{\begin{eqnarray}}
\def\ea{\end{eqnarray}}

\def\C60{A$_x$C$_{60}$}

\def\HgCu3{HgCa$_2$Cu$_3$O$_{8+y}$}
\def\HgCu4{HgBa$_2$Ca$_3$Cu$_4$O$_{10+y}$}
\def\TlCu{Tl$_2$Ba$_2$CuO$_{6+\delta}$}
\def\TlCu3{Tl$_2$Ba$_2$Ca$_2$Cu$_3$O$_{10+y}$}
\def\TlCu4{Tl$_2$Ba$_2$Ca$_3$Cu$_4$O$_{12+y}$}

\def\BiCu3{Bi$_2$Sr$_2$Ca$_{2}$Cu$_3$O$_y$}

\def\C60{A$_x$C$_{60}$}

\begin{document}

\title{Mechanism of High Temperature Superconductivity in a striped
Hubbard Model}
\author{E. Arrigoni}
\affiliation{ Institut f\"ur  Theoretische Physik,
Technische Universit\"at  Graz, Petersgasse 16, A-8010 Graz, Austria}
\affiliation{Institut f\"ur Theoretische Physik und Astrophysik, 
Universit\"at W\"urzburg, am Hubland, 
97074 W\"urzburg, Germany}
\author{E. Fradkin}
\affiliation{Department of Physics, 
University of Illinois at Urbana-Champaign, 1110 W. Green St., 
Urbana, IL 61801-3080}
\author{S.\ A.~Kivelson}
\affiliation{Department of Physics, 
University of California at Los Angeles, 405 Hilgard Ave., 
Los Angeles, CA 90095}

\date{\today}
\begin{abstract}
It is shown, using asymptotically exact methods, that the two dimensional 
repulsive Hubbard
model with strongly modulated interactions exhibits ``high temperature
superconductivity".  Specifically, the explicit modulation, which has the same
symmetry as  period 4
bond-centered stripes, breaks the system into an alternating array of more and 
less
heavily hole doped, nearly decoupled two-leg ladders.  It is shown that this 
system exhibits
a pairing scale determined by the spin-gap of the undoped two-leg ladder, and a 
phase
ordering temperature proportional to a low positive power of the inter-ladder 
coupling.
\end{abstract}

\maketitle

Much has been written concerning the mechanism of high temperature 
superconductivity 
(HTC) since the discovery~\cite{bn} of the cuprate superconductors in
1986, and indeed even before  that.  However,  what is meant by ``the
mechanism'' is rarely defined, and  clearly
evokes   different images for different authors.  
The BCS mechanism, in which pairing 
is a 
consequence of a weak induced {\it attraction} produced by the exchange 
of phonons 
between well defined quasiparticles, is not only consistent with a remarkable 
number of 
experimental facts in conventional superconductors, it is also of well 
established 
theoretical 
validity in simple models.  Because it is a weak coupling theory, 
even the mean-field 
estimate of $T_c$ (which is exponentially small, $T_c \propto \exp(-1/g)$ 
where 
$g$ is 
the induced attraction) is known to be quantitatively reliable~\cite{brout}.   
However, there
are many well known reasons to believe that the BCS
mechanism always leads to low 
$T_c$'s as recently reviewed in Ref.~\onlinecite{review}.  

An alternative idea,
 which has been the focus of much of the theoretical effort
in the field, is that in a doped Mott insulator, high temperature 
superconductivity arises directly from the repulsive interactions between electrons.  
However, even 
as a  point
of principle, the validity of a mechanism of this sort has not been well 
established for any
simple model.  

In this note we demonstrate the existence of a ``high temperature 
superconducting"
phase
of the Hubbard and $t-J$ models on a square lattice with periodically modulated 
parameters \cite{troyer95}  -- 
see Eq. (\ref{model}).
In particular, we show
that  a period 2 modulation can  produce superconductivity with a relatively low $T_c$ in a
restricted doping range, while
a period 4 modulation  produces higher critical temperatures on a broader range of doping.
Specifically, we consider a caricature of a stripe 
ordered
state  consisting of a quasi-one dimensional array of two leg Hubbard ladders 
weakly
coupled to  each other with a hopping matrix element $\delta t$.  For a range of electron densities per site, 
$<n> \equiv 1-x$,  it
has been well  
established~\cite{white,troyer,jeckleman,dagotto,balents,zachar} that
the two leg ladder exhibits a Luther- Emery liquid~\cite{lutheremery} 
phase, with a 
large
spin-gap,
$\Delta_s
\sim J/2$, and a  divergent superconducting susceptibility for 
$T \ll \Delta_s$,
\be
\chi_{SC}(T) \sim \Delta_s\ /\ T^{2-K^{-1}}, 
\label{sc}
\ee
where $K$ is the charge Luttinger parameter, and $T$ is 
the temperature. This sounds like a promising start.  However,  
a non-zero $T_c$ 
is 
impossible in one dimension (1D), so to have a chance of a high 
transition temperature, 
inter-ladder couplings 
must be taken into account.   If all the ladders are equivalent 
(a caricature of 
a period 2 
stripe ordered  or column state~\cite{subir,subir2}),  we shall see that 
this coupling
leads to a superconducting state in a restricted range of small $x$ with rather low $T_c$.
For
more substantial values of $x$, it inevitably leads to  an insulating, 
incommensurate charge
density wave (CDW) state with (in units  in which
the lattice constant is $a=1$) an ordering wave number 
$P=
2\pi  x$.  (It is customary to call this
the 
$4k_F$-CDW since, despite the fact that there is a spin-gap and hence no Fermi 
surface 
whatsoever,  $P=4k_F$, where $k_F$ is the Fermi momentum of a 
one-dimensional non-interacting electron gas at the same electron density.)  That the superconducting 
transition is so easily preempted by CDW order follows from the fact that the CDW  susceptibility 
of the  Luther-Emery liquid diverges as 
\be
\chi_{CDW}(P,T) \sim \Delta_s \ / \ T^{2-K}.
\label{cdw}
\ee
Under most circumstances for repulsive interactions $K < 1$, and hence 
$\chi_{CDW}$ of Eq. (\ref{cdw}) is more strongly divergent than $\chi_{SC}$ 
of Eq. (\ref{sc}).  However,  if we consider 
an alternating array of A and B type ladders (with different 
electron affinities) then the tendency to CDW order is greatly suppressed due to the 
mismatch  between ordering vectors, $P_A$ and $P_B$, on neighboring ladders ~\cite{nature,sll}.  
We shall show that, so long as the exponent inequalities
\be
2 > K_A^{-1}+K_B^{-1} - K_A; \ \ \ 2 > K_A^{-1}+K_B^{-1} - K_B
\label{unequal}
\ee
are satisfied, the  superconducting instability wins out.  
(If the Luttinger parameter is
the  same for 
both ladders, these inequalities reduce to 
$K > K_c\equiv (\sqrt{3}-1)\approx 
0.8$.)  

Under
these  circumstances, the superconducting (Kosterlitz-Thouless) transition 
temperature can be
reliably  estimated
by  treating the 1D fluctuations exactly but the inter-ladder Josephson 
coupling 
${\cal J}$ in mean-
field approximation~\cite{ferrel,erica}.  
\be
T_c \sim \Delta_s \left(\frac {\cal J} {\widetilde W} \right)^{\alpha}; \   
\alpha=\frac {2K_AK_B}{[4K_AK_B-
K_A-K_B]}
\label{Tc}
\ee
where ${\cal J}$ is an effective coupling and $\widetilde W$ is a microscopic 
energy which we will discuss in
detail below;  typically, we find ${\cal J}\sim \delta t^2/J$ and
$\widetilde W\sim J$.  Although
$T_c$ is small  for small ${\cal J}$, it is only power law small. In fact typically
$\alpha  \sim 1$. Because of the mean-field character of this estimate for $T_c$, one expects this to be an upper bound to the actual $T_c$. One also generally expects $T_c$ to be somewhat suppressed by phase fluctuations but typically by no more than a factor of 2. Indeed, a perturbative renormalization-group treatment for small ${\cal J}$
yields the same power law dependence as Eq.~\ref{Tc},  suggesting that
this expression  is asymptotically exact for ${\cal J}<<\widetilde W$.
This fact is supported in Appendix \ref{app:appendix}, where the
accuracy of {\it interchain} mean-field estimates is discussed for
related models.

Since we expect $T_c$ to be smooth function of $\delta
t/{\cal J}$, it is reasonable to extrapolate Eq. \eqref{Tc} to the case in which $\delta t$ is a substantial fraction of ${\cal J}$. This suggests a maximum $T_c$ of order $\Delta_s$, and so can easily account for relatively high transition 
temperatures~\cite{interlayer,rvb}.  This is 
in contrast to the case of an exponentially small $T_c$ as obtained, for example, in a BCS-like mechanism.

\section{ The  striped Hubbard model}  

While the results obtained in this paper are quite 
robust in
the sense that they apply for a broad range of microscopic interactions, to 
establish
their validity it is useful to consider an explicit model.  
The model we study is 
the striped
Hubbard model:
\ba
H=&&-\sum_{<\vec r,\vec r^{\prime>},\sigma} t_{\vec r,\vec r^{\prime}}
[c^{\dagger}_{\vec r,\sigma}c_{\vec r^{\prime},\sigma}+{\rm h.c.}] \\
\label{model}
+&&\sum_{\vec r,\sigma} [\epsilon_{\vec r}
c^{\dagger}_{\vec r,\sigma}c_{\vec r,\sigma}+(U/2)c^{\dagger}_{\vec
r,\sigma}c^{\dagger}_{\vec r,-\sigma}c_{\vec r,-\sigma}c_{\vec r,\sigma}]
\nonumber
\ea
where $<\vec r,\vec r^{\prime}>$ designates nearest-neighbor sites, 
$c^{\dagger}_{\vec r,\sigma}$ creates an  electron on
site $\vec r$ with spin polarization $\sigma=\pm 1$ and satisfies canonical
anticommutation  relations, and
$U>0$ is the repulsion between two electrons on the same site.  
In the limit of 
strong
repulsions,
$U
\gg t_{\vec r,\vec r^{\prime}}$, this model reduces approximately to the 
corresponding $t-J$
model, which operates in the subspace of no double occupied sites, but with an 
exchange
coupling,
$J_{\vec r,\vec r^{\prime}}=4 |t_{\vec r,\vec r^{\prime}}|^2/U$ between 
neighboring spins. 
Our results only
depend on the low-energy physics
of the ladder and, thus,
apply equally to the $t-J$ and Hubbard models.

In the translationally invariant Hubbard model, 
$t_{\vec r,\vec r^{\prime}}=t$ and
$\epsilon_{\vec r}=0$.   The striped 
version of
this model is still translationally invariant along the 
stripe direction
(which we take to 
be the $y$
axis), so $t_{\vec r,\vec r+\hat y}=t$.  However,   
perpendicular to the stripes 
the
hopping matrix takes on alternately large and small values: 
$t_{\vec r,\vec r+\hat 
x} =
t^{\prime}$ for
$r_x=$ even, and $t_{\vec r,\vec r+\hat x} = \delta t\ll t^{\prime}
\sim t
$ for 
$r_x=$ 
odd.  
This defines a ``period 2 striped Hubbard model,'' as shown in Fig. 1.  
For the ``period 4 striped Hubbard model,'' we include a modulated site energy, 
$\epsilon_{\vec r}=
\sqrt{2}\epsilon\cos[\pi r_x/2 -\pi/4]$, which has site energy $\epsilon$ 
and
$-\epsilon$ respectively on every other 2-leg ladder,
with $\epsilon \gg \delta t$.

\begin{figure}[bht]
\begin{center}
\includegraphics[width=0.35\textwidth]{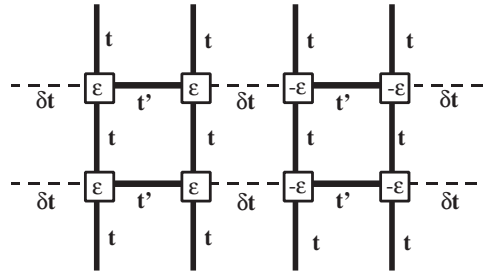} 
\end{center}
\caption
{Schematic representation of the striped Hubbard model analyzed in this paper.}
\end{figure}

\section{ Isolated 2-leg ladder}  

For $\delta t=0$, the model 
breaks up into 
a series
of disconnected 2-leg ladders.  Considerable analytic and numerical effort has 
gone into
studying the properties of  2-leg $t-J$ and Hubbard ladders, and much is known 
about 
them. 
For $x=0$, the undoped two leg ladder has a unique, fully gapped state, 
referred 
to as C0S0
in the notation of Ref. ~\onlinecite{balents}, meaning 0 gapless charge and 0 gapless spin modes.  
In the 
large $U$
limit, the magnitude of the spin-gap of the undoped~\cite{noack,jeckleman} 
ladder is $\Delta_s\approx 
J/2$. Then,
for a substantial range of
$x$  ($0 < x < x_c$) the ladder exhibits a Luther-Emery  or C1S0 phase, with a 
spin-gap that
drops smoothly~\cite{missing} with increasing $x$, and vanishes at a critical 
value of the 
doping,
$x=x_c$.  (This particular Luther-Emery liquid is
known~\cite{jeckleman,troyer,white,dagotto,balents} to have ``d-wave-like" 
superconducting
correlations, in the sense that the pair-field operator has opposite signs 
along the
edge of the ladder (y direction) and on the rungs (x-direction).)  
For $x > x_c$, the
numerical results are scarce, nor is there  uniform agreement concerning 
the number of
phases;  there may~\cite{balents} or may  not~\cite{zachar} be narrow 
ranges of C2S1 and C2S2
phases for $x$ slightly larger than
$x_c$.  At  any rate,
for $x$ large enough, $x_c \le x_c^{\prime} < x < 1$, the ladder manifestly 
enters 
a
Luttinger liquid C1S1 phase, and finally, a trivial C0S0 phase 
when $x=1$ ($<n>=0$).

For the purposes of the present paper, we will confine ourselves to the range 
of parameters where both A and B type ladders are in the Luther-Emery phase. 
The low energy physics (at all energies less than $\Delta_s$) of the two-leg
ladder in the Luther-Emery phase is contained in the free bosonic Hamiltonian 
for 
the
collective charge degrees of freedom,
\be
H=\int dy \frac {v_c} 2\left[K(\partial_y\theta)^2+ \frac {1} {K} 
(\partial_y\phi)^2
\right] +\ldots
\label{Heff}
\ee 
where  $\phi$ is the CDW phase and $\theta$ is the superconducting phase;  
these 
two fields
are dual to each other, and so  satisfy the canonical
commutation relations,
$[\phi(y^{\prime}),\partial_y\theta(y)] = i\delta(y-y^{\prime})$.  
This effective Hamiltonian is general and physical;
the precise $x$ dependence of the
spin-gap, $\Delta_s$, the  charge Luttinger exponent, $K$, the charge velocity, 
$v_c$, and the
chemical potential,
$\mu(x)$, depends on details such as the values of $U/t$ and 
$t^{\prime}/t$.  For
certain cases~\cite{jeckleman,troyer,white} these have been accurately 
computed in
Monte-Carlo studies, and these studies could be straightforwardly extended to 
other values of
the parameters.  

The ellipsis in Eq. (\ref{Heff}) represent cosine potentials, which we will 
not explicitly exhibit here, that produce the Mott gap $\Delta _M$ at
$x=0$.  A consequence of these terms   is that for $x\to 0$, the elementary 
excitations
are charge $2e$ solitons that can either be viewed as spinless Fermions or 
hard-core
bosons, with a dispersion relation $E(k)= \Delta_{M}+\tilde t k^2$.  
One consequence of
this is that~\cite{shulz,white} $K\to 2$ and $v_c\to 2 \pi\tilde t x$ as 
$x\to 0$.
A second consequence is that the renormalized
harmonic theory, which retains only the explicitly exhibited terms in 
Eq. (\ref{Heff}), is
valid in a range of energies which is small in proportion to the effective 
Fermi
energy, $\tilde E_F^{(1D)} = 2\pi\tilde t x^2$.  (An estimate of $\tilde
t \approx t/2$ can be obtained from the DMRG study of the $t-J$ ladder with
$J/t=1/3$ in Ref. ~\onlinecite{troyer}.)

For larger $x$, the numerical studies~\cite{white,troyer,erica} generally 
find that both $K$
and
$\Delta_s$ drop monotonically with increasing $x$.  By the time 
$x=x_1\approx 0.1$, $K$ is
generally found to be close to 1, and by $x=x_c\approx 0.3$, $\Delta_s$ has 
dropped to
values that are indistinguishable from 0, and $K\approx 0.5$.  Thus, over most 
of the
entire Luther-Emery phase, both the SC and the CDW susceptibilities are 
divergent. 
However,  the SC susceptibility is the more divergent only at rather small 
values of $x < x_1$.

Although the charge fields exhaust the low energy degrees of freedom of the Luther-Emery
liquid, when we come to consider the effects
of the single-particle hopping perturbation with small coupling constant $\delta t$, 
we need to 
consider (as virtual
intermediate states) high energy states with the quantum numbers of an 
electron.  Thus,
we need to reintroduce gapped fields $\phi_s$ to represent the spin-degrees 
of freedom.
Since this is standard~\cite{review}, we will not belabor the point; the 
appropriate
continuum fermionic fields are
\be
\Psi_{\pm,\sigma}^{\dagger}\sim
\exp\left\{\sqrt{\pi/2}
\left[\theta\pm\phi+\sigma\theta_s\pm\sigma\phi_s\right]\pm i
Py/2\right\}
\label{psi}
\ee 
where
$\pm$ refer to left and right going fermions with momentum near $\pm P/2$,
respectively, $\sigma=\pm 1$ represents the spin polarization.
It is important to stress that for strongly interacting problems, such as 
the present one,
there is no simple relation between the original lattice fermions and the 
continuum fermion
fields which describe the ``physical ''$\Psi$-fermions of Eq. (\ref{psi}).  
In particular, what appears as a
$2k_F$ CDW expressed in terms of $\Psi$-fermions, would be considered a
$4k_F$ CDW in terms of the original, lattice Fermions.
In terms of these $\Psi$-fields, the component of the charge density operator which
varies with wave  numbers near
$P$ is
\ba
\hat\rho_{P}(y)=\sum_{\sigma}\Psi_{L,\sigma}^{\dagger}\Psi_{R,\sigma}
\propto  \exp[iPy+i\sqrt{2\pi}\phi(y)]
\ea
while the singlet pair creation operator,
\be
\hat\Phi(y)=[\Psi_{L,\uparrow}^{\dagger}\Psi_{R,\downarrow}^{\dagger}+
\Psi_{R,\uparrow}^{\dagger}\Psi_{L,\downarrow}^{\dagger}]
\propto
\exp[i\sqrt{2\pi} \theta],
\label{Phi}
\ee
where in the right-most expressions we have again suppressed the dependence on the spin fields.

Before leaving the single ladder problem, it is worth mentioning a useful 
intuitive 
caricature of its electronic properties.  We picture a singlet pair of 
electrons on
neighboring sites as being a hard-core bosonic ``dimer."  
The undoped ladder can be thought of as a Mott
insulating state of these dimers, with one dimer per rung of the ladder,
{\it i.\ e.\/} a ``valence bond crystal" with lattice spacing one. 
To remove one
electron from the system, we need to destroy one dimer and remove one 
electron, leaving behind a single
electron with spin 1/2 and charge $e$.  However, when we remove a second 
electron from the system, we have
the choice of either breaking another dimer, thus producing two quasiparticles 
with the quantum
numbers of an electron, or of removing the unpaired electron left behind 
by the first removal, thus
producing a new boson - a missing dimer - with charge
$2e$ and spin 0.
The persistence of the spin-gap upon doping the ladder can thus be interpreted 
as
implying that the energy needed to break a dimer  (of order $\Delta_s$) 
is sufficiently large that one
charge $2e$ boson costs less than two charge $e$ quasiparticles.  
At finite $x$, the missing dimers
can be treated as a dilute gas of hardcore bosons. 
That the elementary excitations of the undoped ladder can be constructed in this simple manner
reflects the fact that this is a confining phase~\cite{fk88,read-sachdev}, not a spin liquid.

\section{Inter-ladder interactions}  

We now address the effect of a small, but non-zero
coupling 
({\it i. e.\/} single-particle hopping)
between ladders, $\delta t >0$.  Because of the spin-gap, 
$\delta t$ is  an
irrelevant perturbation in the renormalization group sense, and so 
does not 
directly affect
the thermodynamic state of the system.  However, second order processes 
result in
various induced interactions between neighboring ladders.  These consist
of marginal forward scattering interactions, which are negligible for small
$\delta t$, and potentially relevant Josephson tunnelling and back-scattering
density-density interactions.  

The important (possibly relevant) low energy pieces of these latter interactions are 
most
naturally expressed in terms of the bosonic collective variables defined 
above:
\ba
H^{\prime}= &&-\sum_{j} \int dy \left\{{\cal J} 
\cos[\sqrt{2\pi}(\theta_j-\theta_{j+1})]
\right . 
\\ &&\left .+ {\cal V}\cos[(P_j-P_{j+1})y+\sqrt{2\pi}(\phi_j-\phi_{j+1})]
\right\},
\nonumber
\ea
where  $P_j=2\pi x_j$, with $x_j$ the concentration
of doped holes on ladder $j$, and $\phi_j$ and $\theta_j$ are 
the charge field and its
dual on each ladder.   
Here, again, the form of the low energy interactions
between two Luther Emery liquids  is entirely determined by 
symmetry considerations, but
the magnitude of the  Josephson coupling 
${\cal J}$  and
the induced interaction between CDW's, ${\cal V}$, 
must be computed from microscopics;  
they
are renormalized parameters which result from ``integrating" 
out the high energy
degrees of freedom with energies between the bandwidth 
$W \sim 4t$ 
and 
the renormalized cutoff, $\Delta_s$, or with
wavelengths between
$a$ and
$\xi_s\equiv v_s/\Delta_s$ where $v_s$ is the spin-wave velocity.  
Thus, the
dimensionless measure of the inter-ladder couplings, which for 
instance enter the expressions for $T_c$, are
${\cal J}/\widetilde W$ and ${\cal V}/\widetilde W$ where $\widetilde W = 
\Delta_s/\xi_s$. (So long as $x$ is not too
near $x_c$, $\Delta_s \sim J$, and hence $\widetilde W \sim J$.) 

Quantitative estimates of
${\cal J}$ and
${\cal V}$ could certainly be obtained, given the state of DMRG 
calculations, from studies of
four-leg ladders consisting of two weakly coupled 2-leg 
ladders~\cite{arrigoni}.  However, such
calculations have not, yet, been carried out. 
Fortunately, our qualitative conclusions are not very sensitive to the
values of ${\cal V}$ and ${\cal J}$, which can anyway be estimated 
with reasonable
accuracy from bosonization,  as discussed in Ref.~\onlinecite{erica}.  
The subtlety here is that the inter-ladder hopping 
is expressed in terms of
microscopic
lattice fermions, whereas our low energy theory is expressed in terms of the
$\Psi$-fermions of Eq. (\ref{psi}).  However, since these have the 
same quantum numbers as an electron, and operate on
the scale of
$\Delta_s$, which is large
with respect to $\delta t$,
 there is no reason to expect any large renormalization of the
hopping parameters.  
If we assume that the inter-ladder hopping 
can be approximated as
$\delta t$ times 
an operator representing
the hopping amplitude for  $\Psi$-fermions, then 
from second order perturbation theory  we obtain 
\be 
{\cal J} \approx {\cal V} \sim  A\  (\delta t)^2 / J
\ee
where $A$ is the dimensionless function of $\Delta_s/J$ 
\ba
A= J\int \! dy d\tau \!\left|\langle e^{i\sqrt{\pi/2}[
[\theta_s({\bf r})-\theta_s({\bf 0}) +\phi_s({\bf r})+
\phi_s({\bf 0})]}\rangle_s\right|^2 
\nonumber
\ea
${\bf r}=(y,\tau)$ and $\tau$ denotes imaginary time, 
the expectation value is taken with respect to the spin-fields on the 
decoupled
ladders, and in deriving this expression we have assumed that the charge 
fields are
slowly varying compared to the spin-fields.  Simple scaling arguments 
of the sort
 discussed in Ref. \onlinecite{erica} suggest that 
 $A \sim 1$ as $\Delta_s/J\to 0$. (For further discussion see footnote \onlinecite{otherbands}.) In any
case, so long as  $x$ is not too close to $x_c$,
$\Delta_s$ is of order of the exchange coupling, $J$, so it is reasonable 
to assume $A \sim 1$. 
The only aspects of this estimate which matter qualitatively for our present
purposes is that the two couplings are comparable in size, 
${\cal J}\sim {\cal V}$ and both are small in proportion to $\delta t^2$.  

\section{ Renormalization-group analysis and inter-ladder mean field
  theory}  
The effect of these inter-chain couplings can be
deduced from an analysis of the  lowest order perturbative renormalization 
group equations
in powers of the couplings ${\cal V}$ and ${\cal J}$.  However, 
{\it equivalent} results are
obtained from inter-ladder mean-field theory~\cite{erica,ferrel}, 
which is conceptually
simpler.  These equations are the analogue of the BCS gap equations 
applied to this model,
and are expected to give a quantitatively accurate estimate of $T_c$ 
for small
$\delta t/\Delta_s$
for precisely the same reason.  
A discussion of the accuracy of {\it interchain} mean-field theory is
given in 
Appendix \ref{app:appendix}. 
In the present 
two-dimensional system, $T_c$
should be interpreted as
the onset of quasi-long range order, {\it i. e.} as a
Kosterlitz-Thouless transition.

To implement this mean-field theory, we 
need to 
compute the expectation value $M_j(h_j)= \langle
\cos[\sqrt{2\pi}\theta_j]\rangle$
 of the
pair creation operator  on an isolated ladder, where the expectation value is 
taken with
respect to the mean-field Hamiltonian
\be
H_{MF}= H_j - h_j \int dy 
\cos[\sqrt{2\pi}\theta_j]
\ee
in which $H_j$ is the effective Hamiltonian in Eq. (\ref{Heff}) with 
parameters appropriate to
ladder $j$, and $h_j$ represents the mean-field due to the neighboring 
ladders, and so
satisfies the self-consistency condition,
\be
h_j ={\cal J} [M_{j+1}+M_{j-1}].
\ee

The expression for the mean-field transition temperature can be expressed in 
terms of the
corresponding susceptibility,  $\tilde\chi_{SC}^{(j)}=
\partial M_j(h)/\partial h|_{h=0}$,
which is related to the superconducting susceptibility in Eq. (\ref{sc}) 
by a proportionality
constant which depends on the expectation value of the spin-fields.  
In the case in which
all the ladders are equivalent, this yields the implicit relation 
$2{\cal
J}\tilde\chi_{SC}(T_c)=1$.  For an alternating array of $A$ and $B$ 
type ladders, 
the
expression for the superconducting $T_c$ is easily seen to be
\be
(2{\cal J})^2\ \tilde \chi_{SC}^{(A)}(T_c)\ \tilde\chi_{SC}^{(B)}(T_c)\  = 1.
\label{SCMF}
\ee
Notice that in the case in which the $A$ and $B$ type ladders are identical
Eq. (\ref{SCMF}) reduces properly to the expression for equivalent ladders.
The expression for $\chi_{SC}$ from Eq. (\ref{sc})
can be used to invert Eq. (\ref{SCMF}) to obtain the estimate for
$T_c$ given in Eq. (\ref{Tc}).  

The mean-field
equations for the CDW order are obtained similarly.
The expression for the transition temperature for CDW order 
with wave-vector $P$
is
\be
(2{\cal V})^2\ \tilde \chi_{CDW}^{(A)}(P,T_c)\ 
\tilde\chi_{CDW}^{(B)}(P,T_c)\  = 1
\ee
where the notation is the obvious extension of that used in the superconducting
case.  The best ordering vector is that which maximizes $T_c$.  For $P=P_A$,
$\chi_{CDW}^{(A)}(P_A,T)$ diverges with
decreasing temperature as in Eq. (\ref{cdw}), but $\chi_{CDW}^{(B)}(P_A,T)$ 
saturates
to a finite, low temperature value when $T\sim v_c|P_A-P_B|$.  Thus, even if 
$\chi_{CDW}^{(A)}(P_A,T)$ diverges more strongly with decreasing temperature 
than 
$\chi_{SC}^{(A)}$,
there are two divergent susceptibilities in the expression for the 
superconducting
$T_c$, and only one for the CDW $T_c$; so long as the inequalities 
in Eq. (\ref{unequal})
 are satisfied,  the superconducting transition preempts the CDW
transition!
 
\section{  The $x \to 0$ limit}  

Since $K\to 2$ as $x\to 0$, there 
is necessarily a regime of
small $x$ in which the superconducting susceptibility  on the 
isolated ladder is more
divergent than the CDW susceptibility.  Here, in the presence of weak 
inter-ladder coupling,
even the period 2 striped Hubbard model
({\it i. e.\/}  with $\epsilon=0$)
 is superconducting.  However, 
care must be taken in
this limit, since, as mentioned above, the range of energies over which 
$H$ in Eq. (\ref{Heff})
 is applicable vanishes in proportion to $x^2$.  Fortunately, a complementary
treatment of the problem, which takes into account the additional terms, 
the ellipsis in Eq. (\ref{Heff}), 
can be employed in this limit.  The small $x$ problem can be mapped onto a
problem of dilute, hard-core charge $2e$ bosons (with concentration $x$ 
per rung) with an
anisotropic dispersion, $E(\vec k) = \tilde t k_y^2 - {\cal J} \cos[2k_x] $. 
(The 2 reflects the ladder periodicity.)
Consequently, for small $x$,
\be
T_c \approx 2\pi\ \sqrt{2{\cal J}\tilde t} \ x  F(x) \sim |\delta t| \ x 
\label{bc}
\ee
where $F(x) \sim 1/\ \ln\ln(1/x)$ is never far from 1, and the logarithm
reflects~\cite{fisher} the fact $d=2$ is the marginal dimension for Bose
condensation.  (This result is not substantially different for the period
4 striped Hubbard model, so long as $\epsilon$ is not too large.)  There
is a complicated issue of order of limits when both
$\delta t$ and $x$ are small;  roughly, we expect that $T_c$ will be 
determined by whichever
expression, Eq. (\ref{SCMF}) or Eq. (\ref{bc}), gives the higher $T_c$, 
but with the
understanding that $\chi_{SC}$ must be computed taking into account the 
terms represented by the ellipsis in
Eq. (\ref{Heff}) which cause the susceptibility to vanish as $x\to 0$.

The period 2 striped Hubbard or $t-J$ model indeed has a
superconducting phase at small $x$, because this phase is confined to 
rather small
$x\lesssim 0.1$, where $T_c$ is small in proportion to both
$\delta t$ and $ x$.  Moreover, this may still not be enough to establish a 
mechanism of high temperature superconductivity.  The situation looks 
even worse when the
effects of weak disorder ~\cite{giamarchi}
are considered - when the disorder strength is 
greater than the
intra-ladder energy scale $E_F=2\pi \tilde t x^2$,  it is unlikely that 
any sort of
superconducting coherence will survive.  

For an array of alternating ladders, the range of $x$ for which 
superconductivity dominates
is much extended.  This means that the maximum $T_c$ is much greater, 
and the
superconductivity much more robust to disorder\cite{arrigoni} for the period 4 than the 
period 2 striped Hubbard model.

\section{Optimal Degree of Inhomogeneity for Superconductivity}  

Here we have established that in a strongly striped Hubbard model,
superconductivity is produced directly by the repulsive interactions between 
electrons.  The resulting $T_c$ is proportional to a positive power of $\delta t/t$, 
and so rises as the stripe order becomes less strong.  It is thus natural to ask: 
Is  the stripe order introduced in the present paper simply a calculational crutch which 
permits us obtain well controlled results or is inhomogeneity essential to the mechanism 
of high temperature superconductivity, as has been 
suggested~\cite{sudipandme,ekz,review,arrigoni,ivar} in several previous studies?  

The answer to this question turns on the issue of whether or not the uniform Hubbard model, and its strong coupling relative the $t-J$-model, by themselves support high temperature superconductivity. 
This  question has been the focus of much theoretical research since the discovery of superconductivity in the cuprates. To this date this is not a settled issue. Nor is it the purpose of the present paper to review this extensive literature. Variational calculations have been interpreted both as giving evidence in support \cite{sorella} and against \cite{tklee} superconductivity in Hubbard and $t-J$-type models.  There is also
considerable evidence, from several numerical techniques and high temperature expansions, that the canonical $t-J$ and Hubbard models on a square lattice most likely {\it do not} support high temperature superconductivity; instead they show clear evidence for other types of order which compete with superconductivity~\cite{review,leonid}.   

Assuming that the uniform model does not support high-temperature superconductivity, it follows from the arguments given in the previous sections that there is an optimal degree of inhomogeneity (an optimal degree of stripe order) for a strongly correlated system to exhibit superconductivity. Probably this occurs when $\delta t \sim \Delta_s$. An analogous result was established~\cite{arrigoni} recently in the weakly interacting limit of the 4-leg ladder (itself a caricature  of a
single unit cell of the present model). We should also note that there is nothing essential about having period $4$. In fact, the longer the period the more the CDW instability is suppressed and the larger the range of superconductivity.

\section{Relation to superconductivity in the cuprates}   

While the main purpose of the
present paper was to establish, as a point of principle, that the striped Hubbard model
analyzed here  exhibits high temperature superconductivity, a few comments are 
in order concerning the more general implications of the  present results for the 
mechanism of  superconductivity in the cuprates.  

Firstly, the explicit striped inhomogeneities  introduced here are
a caricature of the  spontaneous symmetry breaking in a charge striped phase. 
However, the model possesses a large spin-gap,  and so does
not contain any of the physics of low energy incommensurate spin-fluctuations 
which are the principle experimental signatures to date of stripe correlations in the 
cuprates.  Secondly, although the superconducting state is ``d-wave-like" in the sense that the 
order parameter changes sign under rotations by $\pi/2$, since the striped Hamiltonian
explicitly breaks this symmetry, there is no precise symmetry distinction between 
d-wave and s-wave superconductivity.  
Thirdly, the superconducting state is not even truly
adiabatically connected to the superconducting state observed in the cuprates, 
because the
existence of a spin-gap implies the absence of gapless ``nodal" 
quasiparticles in 
the
superconducting state.  However, the transition between a node-less and nodal $d$-wave-like
state was studied in Ref. [\onlinecite{granath,subir-lifshitz}], where it was found to be a mean field (Lifshitz) transition with relatively little effect on 
$T_c$.  Moreover, using the
same lines of reasoning employed in that article, it is possible to make 
compelling
(although not entirely rigorous) arguments that upon heavier doping, the 
present  model, too, will exhibit a nodal superconducting state.  We are currently working to 
obtain a more complete treatment of the phase diagram of the present model.

The present  model
realizes the idea that the pairing scale, in this case the spin-gap, can be 
inherited from a parent Mott insulating state.  Moreover, like the underdoped cuprates, 
the gap  scale in the present model is a decreasing function of increasing $x$, while the 
actual superconducting transition occurs at a $T_c$ much smaller than $\Delta_s/2$, 
and 
is
determined by the phase ordering temperature rather than the pairing scale.  
Hence, for $x$ not too close to $x_c$,  this model exhibits a pseudogap regime for
temperatures between
$T_c$ and $T^*\sim \Delta_s/2$, 
reminiscent of that
seen in underdoped cuprates.  However, $T_c$ is always bounded from above by $\Delta_s$ and so tends
to zero as $x\to  x_c$.

\begin{acknowledgments}  
This work was supported, in part, 
by NSF grants No. DMR 01-10329 
at UCLA (SAK), and DMR-01-32990 at the University of Illinois (EF), 
and by a Heisenberg grant (AR 324/3-1) form the DFG (EA).
\end{acknowledgments}

\appendix

\section{Accuracy of the interchain mean-field theory estimates}
\label{app:appendix}

In this section, we discuss the accuracy of the interchain mean-field
theory.  Although no general proof exists (to the best of our knowledge),
we believe that it is asymptotically exact in the present case, at
least to logarithmic accuracy (as defined in Eq. (\ref{logac}) below).  The latter conclusion also
follows, as mentioned in the text, by comparison with 
perturbative RG calculations.  

Quite generally, using an argument based on Griffiths inequalities, one knows that the exact $T_c$ of a general anisotropic ferromagnetic system (not necessarily an Ising model) will obey the bounds: $T_c(J_x)\leq T_c \leq T_c(J_y)$, for $0<J_x\leq J_y$, where $T_c(J)$ is the $T_c$ of an isotropic system of coupling constant $J$. However, for specific systems it is possible to establish more precise estimates of $T_c$. 

As our first example, consider the 2D anisotropic Ising model on a
square lattice, with couplings $J_x$ and $J_y\le J_x$ in the x and y
directions, respectively.  In particular, for the case of the 2D Ising model, it is also known that the exact $T_c$ 
is the solution to the equation\cite{mattis}
\begin{equation}
\sinh(2J_x/T_c)\sinh(2J_y/T_c) =1.
\end{equation}
Interchain mean-field theory for the same model gives the familiar
expression for the mean-field transition temperature, $T_0$:
\begin{equation}
2J_y\chi_{1D}(T_0)=1
\end{equation}
which is analogous to Eq.\eqref{SCMF}, and
 where
\begin{equation}
\chi_{1D}(T)=T^{-1}\exp[2J_x/T]
\label{eq:mfchi}
\end{equation}
is the susceptibility of the 1D Ising model.  In the limit of small
$J_y/J_x$, it thus follows that the ratio
\begin{equation}
\frac {T_0} {T_c} = 1 + \frac {\ln{2}} {\ln[J_x/J_y]} + \ldots
\end{equation}
tends to 1 as $J_y/J_x \to 0$, {\it i.e.\/}  the interchain
mean-field theory is asymptotically exact without any apologies.

Before leaving the Ising example, it is interesting to see how well the
interchain mean-field theory works when extrapolated to the isotropic
case $J_x=J_y=J$.  It is easy to verify that $T_0=3.53 J$ and
$T_c=2J/\ln[1+\sqrt{2}]$, so
\begin{equation}
T_0/T_c = 1.55 \ \ \ {\rm for}\ J_y/J_x=1.
\end{equation}
In general, $T_0/T_c$ rises monotonically from 1 for increasing 
$J_y/J_x$, but $T_0$ gives a reasonably good estimate of $T_c$ over
the entire range of parameters.  (Note, ordinary mean field theory gives
$T_0^{\prime}=2[J_x+J_y]$, which is not much worse than interchain
mean-field theory in the isotropic limit, but $T_0^{\prime}/T_c
\to\infty$ as $J_y/J_x\to 0$.)

Now, we move to the 2D classical XY model on a square lattice - the case
of most direct relevance to the estimates of $T_c$ made in the text.  The
susceptibility of an isolated chain can easily be seen to be
\begin{equation}
\chi_{1D}(T) =\frac 1 {2T} \left(\frac{ I_0(J_x/T) + I_1(J_x/T)}{I_0(J_x/T) -
I_1(J_x/T)}\right)
\label{eq:chixy}
\end{equation}
where $I_n(x)$ is a Bessel function.  
For
$J_y/J_x \ll 1$, Eq. \eqref{eq:mfchi} and Eq.\eqref{eq:chixy}  yield the following estimate of the critical temperature
\begin{equation}
T_0=2\sqrt{J_xJ_y}\left[ 1 + {\cal O}\left(\sqrt{J_y/J_x}\right)\right],
\end{equation}
while $T_0=1.755 J$ in the isotropic limit.  

Unlike the Ising case, no exact results exist for the 2D XY
model.  Extensive Monte-Carlo work has been done on the isotropic 2D XY
model, form which we know\cite{gupta} that the Kosterlitz-Thouless
transition occurs at $T_c=0.89J$, so in this limit
$T_0/T_c\approx 2$.  In the limit of extreme anisotropy, the 2D classical
XY model can be mapped onto the familiar 1D quantum XY (rotor) model with Hamiltonian
\begin{equation}
H=\sum_n \left[ \frac{L_n^2}{2} -\frac{\lambda}{2} \cos(\theta_{n+1}-\theta_n)\right]
\end{equation}
where the coupling constant is $\lambda=2 {J_xJ_y}/{T^2}$ (see Ref. [\onlinecite{fs}])
  The critical value of the coupling of the
quantum rotor model, $\lambda_c$, has been computed quite accurately using a Pad\'e-Borel resummation of the strong-coupling series \cite{hks,hk}. Using the notation of these  papers,  an accurate estimate for the critical coupling to be $\lambda_c=1.8\pm 0.5$ is obtained.  By carefully
inverting this mapping, we get
\begin{equation}
T_c = A \sqrt{J_xJ_y}\left[ 1 + {\cal O}\left(\sqrt{J_y/J_x}\right)\right]
\end{equation}
where $A=1.05\pm 0.1$.  Thus, we see that
\begin{equation}
T_0/T_c \to (2/A) \ \ \ {\rm as} \ \ J_y/J_x \to 0.
\end{equation}
It seems unlikely that the error bars on $A$ are sufficient to be
consistent with a limit of 1.  The interchain mean-field theory is
therefore found  to be asymptotically exact only to logarithmic
accuracy, {\it i.e.}
\begin{equation}
\frac{\ln T_0 }{\ln T_c} \to 1 \ \ \ {\rm as} \ \ \frac{J_y}{J_x} \to 0.
\label{logac}
\end{equation}
This, we believe, is the generically true of interchain mean-field theory
as applied in the present paper.  None-the-less, in all cases where the
exact answers are known, interchain mean-field theory gives estimates of
$T_c$ that are within a factor of 2 of the exact results. This is
certainly sufficiently accurate for present purposes. 

Finally, it is worth mentioning that the 2D XY model is something of a
worst-case example, because 2D is the lower critical dimension and hence
fluctuation effects are anomalously large.  If we consider an anisotropic
3D XY model with couplings $J_x \ge J_y\ge J_z$ in the
three directions, the mean-field transition temperature can still be
readily computed according to
$2(J_y+J_z)\chi_{1D}(T_0)=1$.  Monte-Carlo results
exist\cite{erica2} for the $T_c$ of layered models, $J_y=J_x\equiv J$ for
various values of $J_z/J_x$.  For instance,   
\begin{eqnarray}
&&T_c=1.1 J     \ \ \ \ \ \ T_0/T_c =  1.60  \ \ \ {\rm for } \ \ 
J_z/J=0.01 \nonumber \\
&&T_c=1.324  J     \ \ \  T_0/T_c =  1.41      \ \ \  {\rm for } \ \ 
J_z/J=0.1 \nonumber \\
&&T_c=2.2  J      \ \ \  \ \ \ T_0/T_c =  1.29      \ \ \  {\rm for } \ \ 
J_z/J=1.0. \nonumber 
\end{eqnarray}
 Clearly, even a very small amount of
interplane coupling can be expected to greatly improve the accuracy of
our $T_c$ estimates. (Interplane mean-field theory, of course, is still more accurate, as
shown in Ref. [\onlinecite{erica2}].)


\begin{thebibliography}{99}
\bibitem{missing}
For a restricted range of $x$, the authors of Ref. [\onlinecite{jeckleman}] 
show numerical evidence
indicating that the spin gap decreases smoothly with increasing $x$. 
We are not aware of any
published studies   that carefully trace the spin gap as a function of 
$x$, and in particular 
ones that accurately determine the critical doping,
$x_c$, at which it vanishes.
\end{thebibliography}
\end{document}